\documentclass[preprint,prd,noshowpacs,nofootinbib]{revtex4}
\usepackage{amssymb}
\usepackage{amsmath}

\begin{document}

\title{Self-similarity, conservation of entropy/bits and the black hole information puzzle
\footnote {Awarded a fourth prize in the 2013 FQXi's Essay Contest ``It From Bit, or Bit From It"} \\}

\author{Douglas Singleton}
\email{dougs@csufresno.edu}
\affiliation{Department of Physics, California State University Fresno, Fresno, CA 93740-8031, USA
and \\
Department of Physics,
Institut Teknologi Bandung, Jalan Ganesha 10 Bandung 40132, Indonesia}

\author{Elias C. Vagenas}
\email{elias.vagenas@ku.edu.kw}
\affiliation{Theoretical Physics Group, Department of Physics, Kuwait University, 
P.O. Box 5969, Safat 13060, Kuwait}

\author{Tao Zhu}
\email{Tao_Zhu@baylor.edu}
\affiliation{GCAP-CASPER, Physics Department, Baylor University, Waco, TX 76798-7316, USA and \\
Institute for Advanced Physics \& Mathematics,
Zhejiang University of Technology, Hangzhou, 310032, China}


\begin{abstract}
John Wheeler coined the phrase ``it from bit" or ``bit from it" in the 1980s.
However, much of the interest in the connection between information, i.e. ``bits",
and physical objects, i.e. ``its", stems from the discovery that black holes have
characteristics of thermodynamic systems having entropies and temperatures.
This insight led to the information loss problem -- what happens to the ``bits" when the
black hole has evaporated away due to the energy loss from Hawking radiation? In this essay
we speculate on a radical answer to this question using the assumption of self-similarity
of quantum correction to the gravitational action and the requirement
that the quantum corrected entropy be well behaved in the limit when the black hole
mass goes to zero.
\end{abstract}

\maketitle
%
%
%
\section{Self-Similarity and order-$\hbar ^n$ Quantum Gravity Corrections}
\par\noindent
In this essay we look at the connection between physical objects, i.e. ``its", and information/entropy,
i.e ``bits",\footnote{There is an equivalence or connection between information, entropy and bits and we will
use these terms somewhat interchangeably throughout this essay. A nice overview of the close relationship
between information, entropy and bits can be found in reference \cite{susskind}.}
in the context of black hole physics. In particular, we focus on the relationship between
the initial information/entropy contained in the horizon of a Schwarzschild black hole and
the final entropy carried by the outgoing, {\it correlated} photons of Hawking radiation. The
correlation of the photons comes from taking into account conservation of energy and the back reaction of the
radiation on the structure of the Schwarzschild space-time in the tunneling picture \cite{wilczek, pad}
of Hawking radiation. Since, in the first approximation, Hawking radiation is thermal there are
no correlations between the outgoing Hawking radiated photons. This leads to the information loss
puzzle of black holes which can be put as follows: The original black hole has an entropy given
by $S_{BH} = \frac{4 \pi k_B G M^2}{c \hbar}$ which can be written as
$S_{BH}= \frac{ k_{B} A }{4 l_{Pl} ^2}$ where $A=4 \pi r_H ^2$ is the horizon area of the black hole
and $r_H = \frac{2 G M}{c^2}$ is the location of the horizon \cite{bekenstein}. One can think of this areal entropy
as being composed of Planck sized area ``bits", $A_{Pl} = l_{Pl} ^2$, where the Planck length is
defined as $l_{Pl} = \sqrt{\frac{\hbar G}{c^3}}$. If Hawking radiation were truly thermal, then
the entropy of the outgoing thermal radiation would be larger than this Bekenstein area entropy. Since
entropy increases, some information is lost. But this violates the prime directive of quantum mechanics
that quantum evolution should be unitary and, thus, information and entropy should be conserved.

To begin our examination of these issues of the thermodynamics of black holes and the loss versus
conservation of information, we lay out our basic framework. We will consider a massless scalar field
$\phi ({\bf x}, t)$ in the background of a Schwarzschild black hole whose metric is given by

\begin{eqnarray}
\label{schwarz}
ds^2=- \left( 1 - \frac{2 M}{r} \right) dt^2+\frac{1}{\left( 1 - \frac{2 M}{r} \right)} dr^2 +r^2d\Omega^2 ~,
\end{eqnarray}

\par\noindent
in units with $G=c=1$. From here onward in the essay we will set
$G=c=1$ but will keep $\hbar$ explicitly. The horizon is located by setting
$1 - \frac{2 M}{r_H}=0$ or $r_H = 2 M$. Into this space-time, we place a massless scalar field
obeying the Klein-Gordon equation

\begin{eqnarray}
\label{KG equation}
-\frac{\hbar^2}{\sqrt{-g}}\partial_\mu(g^{\mu\nu}\sqrt{-g}\partial_\nu)\phi=0 ~.
\end{eqnarray}

\par\noindent
By the radial symmetry of the Schwarzschild space-time as given by Eq. \eqref{schwarz},
the scalar field only depends on $r$ and $t$. Expanding $\phi (r,t)$ in a WKB form gives

\begin{eqnarray}
\label{scalar field WKB}
\phi(r,t)=\exp\left[\frac{i}{\hbar}I(r,t)\right]
\end{eqnarray}

\par\noindent
where $I(r,t)$ is the one-particle action which can be expanded in powers of $\hbar$ via
the general expression

\begin{eqnarray}
\label{expansion of the action}
I(r,t)=I_0(r,t)+\sum_{j=1} ^\infty \hbar^j I_j(r,t).
\end{eqnarray}

\par\noindent
Here, $I_0(r,t)$ is the classical action and $I_j(r,t)$ are order $\hbar ^j$quantum corrections.
We now make the assumption that quantum gravity is {\it self-similar} \footnote{Broadly speaking, self-similarity
means that a system ``looks the same" at different scales. A standard example is the Koch snowflake \cite{koch}
where any small segment of the curve has the same shape as a larger segment. Here, self-similarity is applied
in the sense that as one goes to smaller distance scales/higher energy scales by going to successive orders in $\hbar$
that the form of the quantum corrections remains the same.}
in the following sense: the higher order corrections to the action, $I_j(r,t)$, are proportional
to $I_0(r,t)$, i.e. $I_j(r,t) = \gamma_j I_0 (r, t)$ where $\gamma_j$ are constants.
With this assumption, Eq. \eqref{expansion of the action} becomes

\begin{eqnarray}
\label{expansion-elias}
I(r,t)=\left(1+\sum_{j=1} ^\infty \gamma_j \hbar^j \right)I_0(r,t)~.
\end{eqnarray}

\par\noindent
From Eq. \eqref{expansion-elias}, one sees that $\gamma_j \hbar^j$ is dimensionless. In the units
we are using, i.e. $G=c=1$, $\hbar$ has units of the Planck length squared, i.e. $l _{Pl} ^2$, thus
$\gamma _j$ should have units of an inverse distance squared to the $j^{th}$ power. The natural distance
scale defined by Eq. \eqref{schwarz} is the horizon distance $r_H = 2 M$, thus

\begin{eqnarray}
\label{gammaj}
\gamma_j=\frac{\alpha_j}{r_H^{2j}}
\end{eqnarray}

\par\noindent
with $\alpha_j$ dimensionless constants which we will fix via the {\it requirement}
that information/entropy be well behaved in the $M \rightarrow 0$ limit. Thus, in this
way we will obtain an explicit, all orders in $\hbar$ correction to the entropy and
show how this gives a potential solution to the black hole information puzzle.\\
%
%
%
\section{Black hole entropy to all orders in $\hbar$}
\par\noindent
In \cite{jhep} the set-up of the previous section was used to obtain an expression for the
quantum corrected temperature of Hawking radiation \cite{hawking} to
all orders in $\hbar$. This was done by applying the tunneling method introduced in
\cite{wilczek, pad} to the WKB-like expression given by Eqs. \eqref{scalar field WKB},
\eqref{expansion-elias}, and \eqref{gammaj}. From \cite{jhep}, the quantum corrected
Hawking temperature is given as

\begin{eqnarray}
\label{corrected Hawking temperature}
T=\frac{\hbar}{8\pi M}\left(1+\sum_{j=1} ^\infty \frac{\alpha_j\hbar^j}{r_H^{2j}}\right)^{-1}.
\end{eqnarray}

\par\noindent
In this expression, $\frac{\hbar}{8\pi M}$ is
the semi--classical Hawking temperature and the other terms are higher order quantum corrections. At this
point, since the $\alpha _j$'s are completely undetermined, the expression in Eq. \eqref{corrected Hawking temperature}
does not have much physical content but is simply a parameterizing of the quantum corrections. However,
by requiring that the quantum corrected black hole entropy be well behaved in the limit
$M \rightarrow 0$, we will fix $\alpha _j$'s and show how this leads to conservation of information/entropy,
thus providing an answer to the black hole information loss puzzle.

Using Eq. \eqref{corrected Hawking temperature}, we can calculate the Bekenstein entropy to all orders
in $\hbar$. In particular, the Bekenstein entropy of black holes can be obtained by integrating the first law of thermodynamics,
$dM = T dS$ with the temperature $T$ given by Eq. \eqref{corrected Hawking temperature}, i.e. $S = \int \frac{dM}{T}$.
Integrating this over the mass, $M$, of the black hole (and recalling that $r_H = 2 M$) gives the modified entropy 
as a function of $M$

\begin{equation}
\label{entropy-corrected}
S _{BH} (M)  = \frac{4 \pi}{\hbar} M^2 + \pi \alpha _1 \ln \left( \frac{M^2}{\hbar} \right)
- \pi \sum _{j=1} ^\infty \frac{ \alpha_{j+1}}{4^j j} \left( \frac{\hbar}{M^2} \right) ^j.
\end{equation}

\par\noindent
To lowest order $S _0 (M) = \frac{4 \pi}{\hbar} M^2$ for which the limit
$M \rightarrow 0$ is well behaved, i.e. $S_0 (M \rightarrow 0) \rightarrow 0$, as expected
since as the mass vanishes so should the entropy.
On the other hand, for the first, logarithmic correction as well as the other higher
corrections, the quantum corrected entropy diverges. One way to fix these
logarithmic and power divergences in $S _{BH} (M)$ as $M \rightarrow 0$ is to postulate
that the Hawking radiation and resulting evaporation turn off when the black hole reaches some
small, ``remnant" mass $m_R$ \cite{remnant}. Here, we take a different path -- by assuming
that quantum corrected black hole entropy should not diverge in the $M \rightarrow 0$ limit 
we will obtain a condition that fixes almost all the unknown $\alpha_j$'s. To accomplish this,
the third term in Eq. \eqref{entropy-corrected} should sum up to a logarithm which can then be combined
with the second logarithmic term to give a non-divergent entropy, i.e. $S(M \rightarrow 0 ) \ne \pm \infty$.
This condition can be achieved by taking the $\alpha_j$'s as

\begin{equation}
\label{alpha}
\alpha _{j+1} = \alpha _1 (-4)^j ~~ \text{for} ~~ j=1,2,3... ~~.
\end{equation}

\par\noindent
This again shows self-similarity since all the  $\alpha_j$'s are proportional to each other.
For this choice in Eq. \eqref{alpha}, the sum in Eq. \eqref{entropy-corrected}, i.e. the third term,
becomes $+\alpha _1 \pi \ln (1 + \hbar/M^2)$. Combining this term with the second, logarithmic
quantum correction, the entropy takes the form

\begin{equation}
\label{entropy-corrected2}
S _{BH} (M)  = \frac{4 \pi}{\hbar} M^2 + \pi \alpha _1 \ln \left( 1 + \frac{M^2}{\hbar} \right)~.
\end{equation}

\par\noindent
As $M \rightarrow 0$, this ``all orders in $\hbar$" entropy tends to zero, i.e. $S _{BH} (M) \rightarrow 0$.
There is a subtle issue with identifying the sum in Eq. \eqref{entropy-corrected}
with $\alpha _1 \pi \ln (1 + \hbar/ M^2)$ -- strictly this is only valid for $\sqrt{\hbar} < M$, i.e. when the
mass, $M$, is larger than the Planck mass. However, we can use analytic continuation to define the sum via
$\alpha _1 \pi \ln (1 + \hbar/M^2)$ even for $\sqrt{\hbar} > M$. This is analogous to the
trick in String Theory \cite{zwiebach} where the sum $\sum _{j=1} ^\infty j$ is defined as $\zeta (-1) = -\frac{1}{12}$
using analytic continuation of the zeta function, i.e. $\zeta (s) = \sum _{n=1} ^\infty n^{-s}$.
Other works \cite{beyond1} have investigated quantum corrections to the entropy beyond the classical level. These
expressions, in general, involve logarithmic and higher order divergences as $M \rightarrow 0$ as we also
find to be the case for our generic expression in Eq. \eqref{entropy-corrected}. However, here,
as a result of our assumption of self-similarity of the $\hbar ^n$ corrections, we find an expression
for $S _{BH} (M)$ which has a well behaved $M \rightarrow 0$ limit.

This ``lucky" choice of $\alpha_j$'s in Eq. \eqref{alpha} which gave the all orders
in $\hbar$ expression for $S _{BH} (M)$ in Eq. \eqref{entropy-corrected2} was motivated by
making the primary physical requirement that the entropy of the black hole be
well behaved and finite. Usually, the focus in black hole physics is to find some way to tame the
divergent Hawking temperature in the $M \rightarrow 0$ limit whereas here the primary physical
requirement has been on making sure that the entropy/information content of the black hole
is well behaved to all orders in $\hbar$.

The expression for $S _{BH} (M)$ still contains an arbitrary constant, namely $\alpha _1$, which is the first order
quantum correction. This first order correction has been calculated in some theories of quantum gravity.
For example, in Loop Quantum Gravity one finds that $\alpha _1 = - \frac{1}{2}$ \cite{meissner}. Once
$\alpha _1$ is known, our assumption of self-similarity and the requirement that information/entropy
be well behaved fixes the second and higher order quantum corrections.
One can ask how unique is the choice in Eq. \eqref{alpha}? Are there
other choices which would yield $S _{BH} (M=0) \rightarrow 0$? As far as we have been
able to determine, there are no other choices of $\alpha _j$'s that give $S (M=0) \rightarrow 0$,
{\it and} also conserves entropy/information as we will demonstrate in the next section.
However, we have not found a formal proof of the uniqueness of the choice of $\alpha _j$'s.

If one leaves $\alpha _1$ as a free parameter -- does not fix it to the Loop Quantum Gravity
value, i.e. $\alpha _1 = - \frac{1}{2}$ --, then there is an interesting dividing point in the behavior of
the entropy in Eq. \eqref{entropy-corrected2} at $\alpha _1 = -4$. For $\alpha _1 \ge -4$, the entropy
in Eq. \eqref{entropy-corrected2} goes to zero, i.e. $S_{BH} = 0$, only at $M = 0$. For $\alpha _1 < -4$, the entropy
in Eq. \eqref{entropy-corrected2} goes to zero, i.e. $S_{BH} = 0$, at $M = 0$ and also at some other value $M=M^* >0$ where
$M^*$ satisfies the equation $\frac{4 \pi}{\hbar} (M^*)^2 + \pi \alpha _1 \ln \left( 1 + \frac{(M^*)^2}{\hbar} \right) = 0$. Thus,
depending on the first quantum correction $\alpha _1$ the black hole mass can vanish if $\alpha _1 \ge -4$,
or one can be left with a ``remnant" of mass $M^*$ if  $\alpha _1 < -4$. It might appear that one could rule out this
last possibility since for $M^* >0$ the black hole would still have a non-zero temperature via Eq.
\eqref{corrected Hawking temperature} and, thus, the black hole should continue to lose mass via evaporation
leading to masses $M < M^*$ which would give $S<0$ for the case when $\alpha _1 < -4$. However,
if the Universe has a positive cosmological constant, i.e. space-time is de Sitter, then the Universe will be
in a thermal state at the Hawking-Gibbons temperature, i.e. $T_{GH} = \frac{\hbar \sqrt{\Lambda}}{2 \pi}$ \cite{gibbons}
where $\Lambda >0$ is the cosmological constant. Thus, if the quantum corrected black hole temperature from
Eq. \eqref{corrected Hawking temperature} becomes equal to $T_{GH}$ the evaporation process can stop at this
finite temperature and still consistently have $S=0$. This situation would give some interesting and non-trivial
connection between the Universal parameter $\Lambda$ and the final fate of every black hole (in the case
when $\alpha _1 < -4$).\\
%
%
%
\section{Conservation of energy, entropy/information and solution to the information loss puzzle}
\par\noindent
We now want to show that the initial (quantum corrected) entropy of the
black hole given in Eq. \eqref{entropy-corrected2} can be exactly
accounted for by the entropy of the emitted radiation so that entropy/information, i.e. ``bits",  is
conserved. The fact that this happens depends crucially on the specific, logarithmic form of the
quantum corrected entropy in Eq. \eqref{entropy-corrected2}. This, retrospectively, puts
an additional constraint on the $\alpha _j$'s from Eq. \eqref{alpha} -- other choices of $\alpha_j$'s
would not in general lead to both a well behaved $S$ in the $M \rightarrow 0$ limit {\it and}
to entropy/information conservation. As we will see, this conservation of information/entropy is connected with the
conservation of energy.

To start our analysis, we note that in the picture of Hawking radiation as
a tunneling phenomenon the tunneling rate, i.e. $\Gamma$, and the change in entropy are related by \cite{wilczek}

\begin{eqnarray}
\label{entropy change2}
\Gamma =e^{\Delta S _{BH}} ~.
\end{eqnarray}

\par\noindent
When the black hole of mass $M$ emits a quanta of energy $\omega$ energy conservation tells us that
the mass of the black hole is reduced to $M -\omega$. Connected with this, the entropy of the
black hole will change according to $\Delta S _{BH}=S _{BH} (M-\omega)-S _{BH} (M)$ \cite{parikh, vagenas}.
Using Eq. \eqref{entropy-corrected2} for the quantum corrected entropy, one obtains for the change in entropy

\begin{equation}
\label{DeltaS}
\Delta S _{BH} =-\frac{8\pi}{\hbar}\omega \left( M-\frac{\omega}{2} \right)
+ \pi\alpha_1 \ln \left[ \frac{\hbar + (M-\omega)^2}{\hbar +  M ^2} \right].
\end{equation}

\par\noindent
Combining Eqs. \eqref{entropy change2} and \eqref{DeltaS}, the corrected tunneling rate takes the form

\begin{eqnarray}
\label{tunneling}
\Gamma (M; \omega)= \left( \frac{\hbar + (M-\omega)^2}{\hbar +  M ^2}\right)^{\pi\alpha_1}\exp{\left[-\frac{8\pi}{\hbar}\omega
\left(M-\frac{\omega}{2} \right) \right]} .
\end{eqnarray}

\par\noindent
The term $\exp{\left[-\frac{8\pi}{\hbar}\omega \left(M-\frac{\omega}{2} \right) \right]}$
represents the result of energy conservation and back reaction on the tunneling rate \cite{parikh, vagenas};
the term to the power $\pi \alpha_1$ represents the quantum corrections to all orders in $\hbar$.
This result of being able to write the tunneling rate as the product of these two effects, namely
back reaction and quantum corrections, depended crucially on the specific form of $S _{BH} (M)$ and $\Delta S _{BH}$ from
Eqs. \eqref{entropy-corrected2} and \eqref{DeltaS}, respectively, which in turn was crucially tied 
to our specific choice of $\alpha_j$'s in Eq. \eqref{alpha}. Note that even in the classical limit, where 
one ignores the quantum corrections by setting $\pi \alpha_1 =0$, there is a deviation from a thermal spectrum 
due to the $\omega ^2$ term in the exponent in Eq. \eqref{tunneling}.

We now find the connection between the tunneling rate given by Eq. \eqref{tunneling} and the entropy of
the emitted radiation, i.e. $S_{rad}$. Assuming that the black hole mass is completely radiated away,
we have the relationship $M=\omega_1 + \omega_2 +...+ \omega _n = \sum _{j=1} ^n \omega _j$ between the mass of the black hole
and the sum of the energies, i.e. $\omega_j$, of the emitted field quanta. The probability for this radiation to occur is
given by the following product of $\Gamma$'s \cite{information} which is defined in Eq. \eqref{tunneling}

\begin{equation}
\label{probability}
P _{rad} = \Gamma (M; \omega_1) \times \Gamma (M-\omega_1 ; \omega _2) \times ... \times
\Gamma \left( M- \sum _{j=1} ^{n-1} \omega _j ; \omega _n \right)~.
\end{equation}

\par\noindent
The probability of emission of the individual field quanta of energy $\omega _j$ is given by

\begin{eqnarray}
\label{probability2}
\Gamma (M; \omega_1) &=&
\left( \frac{\hbar + (M-\omega_1)^2}{\hbar+ M ^2}\right)^{\pi\alpha_1}\exp{\left[-\frac{8\pi}{\hbar}\omega_1
\left(M-\frac{\omega_1}{2} \right) \right]} ~, \nonumber \\
\Gamma (M-\omega_1; \omega_2) &=& \left( \frac{\hbar +  (M-\omega_1 -\omega_2)^2 }{\hbar +  (M-\omega_1) ^2}\right)^{\pi\alpha_1}
\exp{\left[-\frac{8\pi}{\hbar}\omega_2  \left(M -\omega_1 - \frac{\omega_2}{2} \right) \right]} ~, \nonumber \\
&\,& \nonumber \\
&&..... ~, \\
&\,& \nonumber \\
\Gamma \left( M- \sum _{j=1} ^ {n-1} \omega_j; \omega_n \right) &=&
\left( \frac{\hbar + (M- \sum _{j=1} ^ {n-1} \omega_j -\omega_n)^2}{\hbar +  (M-\sum _{j=1} ^ {n-1} \omega_j ) ^2
}\right)^{\pi\alpha_1}
\exp{\left[-\frac{8\pi}{\hbar}\omega_n  \left(M - \sum _{j=1} ^ {n-1} \omega_j - \frac{\omega_n}{2} \right) \right]} \nonumber \\
&=& \left( \frac{\hbar}{\hbar + (M-\sum _{j=1} ^ {n-1} \omega_j ) ^2 }\right)^{\pi\alpha_1} \exp( -4 \pi  \omega _n ^2 / \hbar) ~.
\nonumber
\end{eqnarray}

\par\noindent
The $\Gamma$'s of the form $\Gamma (M-\omega_1 -\omega_2-...-\omega_{j-1} ; \omega_j)$ represent the probability for the
emission of a field quantum of energy $\omega _j$ with the condition that first the field quanta of energy
$\omega_1 + \omega_2+...+\omega_{j-1}$ have been emitted in sequential order.

Using Eq. \eqref{probability2} in Eq. \eqref{probability}, we find the total probability for the
sequential radiation process described above

\begin{equation}
\label{probability3}
P_{rad} = \left( \frac{\hbar}{\hbar+ M ^2}\right)^{\pi\alpha_1} \exp( -  4\pi M ^2 / \hbar) ~.
\end{equation}

\par\noindent
The black hole mass could also have been radiated away by a different sequence of field quanta energies, e.g.
$\omega_2 +\omega_1+...+\omega_{n-1} + \omega_n$.
Assuming each of these different processes has the same probability, one can count the number of
microstates, i.e. $\Omega$, for the above process as $\Omega = 1/P_{rad}$. Then, using the
Boltzmann definition of entropy as the natural logarithm of the number of microstates, one gets for the
entropy of the emitted radiation

\begin{equation}
\label{rad-entropy}
S_{rad} = \ln (\Omega ) = \ln \left( \frac{1}{P_{rad}} \right) =
\frac{4 \pi}{\hbar} M^2 + \pi \alpha _1 \ln \left( 1 + \frac{M^2}{\hbar} \right) ~.
\end{equation}

\par\noindent
This entropy of the emitted radiation is identical to the original entropy of the
black hole (see Eq. \eqref{entropy-corrected2}), thus entropy/information/``bits" are conserved
between the initial (black hole plus no radiation) and final (no black hole plus radiated field quanta) states.
This implies the same number of microstates between the initial and final states and, thus, unitary evolution.
This then provides a possible resolution of the information paradox when the specific conditions are imposed.

The above arguments work even in the case where one ignores the quantum corrections \cite{information},
i.e. if one lets $\alpha _1 =0$. While interesting, we are not sure how significant this is since almost
certainly quantum corrections will become important as the black mass and entropy go to zero.

In this essay, we have examined the interrelationship of ``bits" (information/entropy) and ``its" (physical
objects/systems) in the context of black hole information. By requiring that the higher order quantum corrections
given in Eq. \eqref{expansion of the action} be self-similar in the sense $I_j (r,t) \propto I_0$, and that the
associated entropy/information of the black hole as given in Eq. \eqref{entropy-corrected} be well behaved in the
limit when the black hole mass goes to zero, we were able to relate all the higher order quantum corrections
as parameterized by the $\alpha_j$'s in terms of the first quantum correction $\alpha _1$. This proportionality
of all $\alpha _j$'s is another level of self-similarity. The final
expression for this quantum corrected entropy, namely Eq. \eqref{entropy-corrected2}, when combined with energy conservation
and the tunneling picture of black hole radiation allow us to show how the original ``bits" of black hole
information encoded in the horizon were transformed into the ``its" of the outgoing correlated Hawking photons,
thus providing a potential all orders in $\hbar$ solution to the black hole information loss puzzle.

Finally, as a last comment, it should be stressed that the assumption that the higher order corrections are self-similar 
in the sense given in Eq. (5) (where we take $I_j  \propto  I_0$) and in Eq. (9) (where we take 
$\alpha _{j+1} \propto \alpha _1$) is not at all what one would expect of the quantum corrections in the canonical 
approach to quantum gravity where the quantum corrections would in general generate any possible terms consistent 
with diffeomorphism-invariance. However, this is the problematic aspect of the canonical approach to quantum gravity 
and, thus, it is worth looking into radical suggestions such as the one proposed here, i.e. that the higher order quantum corrections
are greatly simplified by the assumption of self-similarity. This simplification might be seen as an extreme form 
of the holographic principle of quantum gravity as expounded in [1]. In this monograph, it is pointed out that the 
entropy of a black hole scales with the area of the horizon while for a normal quantum field theory the entropy 
will scale as the volume.  The conclusion of this observation is that ``there are vastly fewer degrees of freedom 
in quantum gravity than in any QFT" (see chapter 11 of [1]). This assumption of self-similarity
of the quantum corrections is in the vein of the holographic principle, since making the assumption of 
self-similarity means there are vastly fewer types/forms that the quantum corrections can take as 
compared to canonical quantum gravity.\\
{\par\noindent {\bf Acknowledgments:}}
%
%
 There are two works -- one on self-similarity 
\cite{piero} and one on the peculiar relationship between long distance/IR scales and 
short distance/UV scales in quantum gravity \cite{euro}  -- which helped inspire parts of this work. 
%


\end{document}